# Interplays between charge and electric field in perovskite solar cells: charge transport, recombination and hysteresis

Jiangjian Shi, Huiyin Zhang, Xin Xu, Dongmei Li, Yanhong Luo, Qingbo Meng*

*Key Laboratory for Renewable Energy, Chinese Academy of Sciences; Beijing Key Laboratory for New Energy Materials and Devices; Institute of Physics, Chinese Academy of Sciences, Beijing 100190, P. R. China*

\* Corresponding author qbmeng@iphy.ac.cn

**Abstract:** Interplays between charge and electric field, which play a critical role in determining the charge transport, recombination, storage and hysteresis in the perovskite solar cell, have been systematically investigated by both electrical transient experiments and theoretical calculations. It is found that the light illumination can increase the carrier concentration in the perovskite absorber, thus enhancing charge recombination and causing the co-existence of high electric field and free carriers. Meanwhile, the cell shows a similar charge storage and junction mechanism to that of the multicrystalline silicon solar cell, where the junction electric field determines the charge collection and distribution. Furthermore, it is demonstrated that the static charge of both the doping and defect coming from ion (vacancy) migration can significantly influence the electric field inside the cell, thus affecting the charge collection and recombination, which could be the origins for the widely-concerned hysteresis behaviors.

**I. Introduction**

As an emerging photoelectric system, the perovskite solar cell has attracted wide interest in the past few years due to its outstanding photovoltaic performance and low-cost solution-based fabrication processes[1-12]. For higher efficiency and solving the widely concerned problems of anomalous photoelectric behaviors and low stability, it is important to establish a fundamental understanding toward the physical and chemical properties of the material and working mechanisms of the cell. Numerous valuable efforts have been paid to clarify the general and specific charge processes in this cell regarding charge generation, separation, transport and recombination from the perspective of time and space.[13-28] Besides the distinctive advantages in low non-radiative recombination, long carrier lifetime and high light absorption, the perovskite organic lead halides are found to have ion migration and dipole polarization, which makes this system more complicated in photoelectric process.[27-28]

The perovskite solar cell was developed from the mesoscopic sensitized solar cell,[1, 2, 4] but shows remarkable photovoltaic performance in the structure of inorganic and organic heterojunction solar cells.[3, 5, 6] Until now, several most fundamental physical issues about this cell have not been addressed, for example, the similarity and difference in charge processes and working mechanisms between this cell and the conventional sensitized or junction-type cells, the role of the heterojunction and electric field in the cell and the influence of ion migration and dipole alignment on the photoelectric characteristics. For a photoelectric device, the most basic parameters are charge, electric field and the interplays between them, which directly determine the microscopic charge transport and recombination processes and thus the macroscopic photoelectric behavior of a solar cell. Clarifying them in the perovskite solar cell would benefit for understanding the fundamental issues mentioned above and the widely-concerned hysteresis and stability problems.

In this work, the complicated but important interplays between charge and electric field in the perovskite solar cell are investigated with the help of both experiments and theoretical calculations. Three aspects including charge recombination, collection and storage, and microscopic charge processes behind the photoelectric hysteresis have been revealed. Firstly, an electro-optical independently modulated transient photovoltage/photocurrent measuring system is designed, where external bias voltages and light illumination are independently introduced to modulate the charge and electric field of the cell, and the interplays between them are depicted with the electrical transients. Secondly, the recombination of the cell is found to be very sensitive to light illumination in the short-circuit condition, implying high concentration of free charge stored inside the perovskite absorber. Thirdly, the charge collection and storage properties of the perovskite solar cell are revealed almost identical to that of the silicon cell, no matter in dark or under illumination. Fourthly, the significant influence of static charge from the doping and defect on the electric field and the influence of electric field on the collection and recombination of free charge have also been demonstrated, which could be the origins for the widely-concerned hysteresis behaviors.

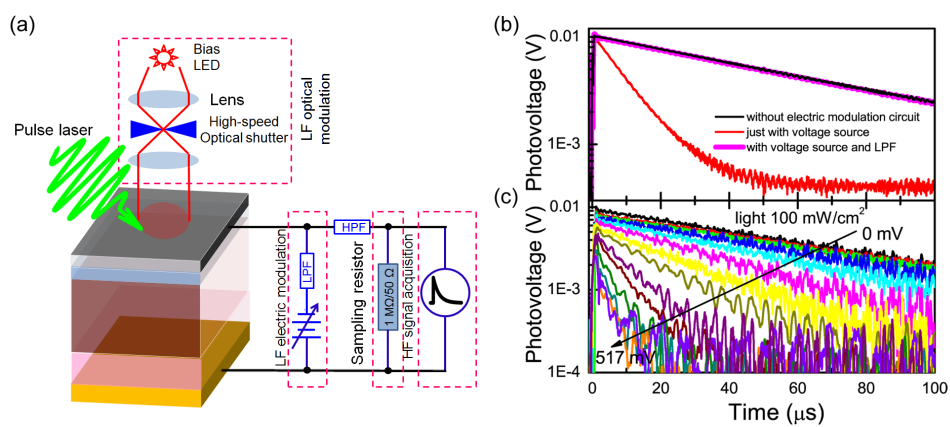

Figure 1. (a) Schematic diagram of the electro-optical independently modulated transient photovoltage/photocurrent measuring system; transient photovoltage results of the multicrystalline silicon solar cell (b) measured with different electrical modulation circuits (dark, 0 V) and (c) under different bias DC bias voltages and a steady-state light illumination of 100 mW/cm$^2$.

## II. Electro-optical independently modulated transient photovoltage/photocurrent system

Transient photocurrent/photovoltage is a general method to investigate charge transport and recombination in a photoelectric system, which has been widely applied in the sensitized and polymer solar cells.[29-31] However, this conventional method can only provide the charge transport or recombination information in the short-circuit or open-circuit conditions,[30] respectively, which cannot expand its application in studying the complicated charge processes of a solar cell in different working conditions, for example, different bias voltages, different light illuminations and the transition photoelectric states between them. This disadvantage can be attributed to the RC and bandwidth limitation of the measuring system and the measured cell. Due to this limitation, few works based on this method has been reported in the study of perovskite solar cell.[33-35] Here, we develop a modulated method to accurately measure the charge transport and recombination properties of a solar cell in practical working conditions (e.g. under different bias voltages and lights) by the introduction of electrical low-pass filter (LPF) and high-pass filter (HPF). Our newly designed measuring system is depicted in Figure 1(a). The transient charge in the cell is excited by a pulse laser (Brio, 532 nm, 4 ns), and the high-frequency photocurrent/photovoltage signal is recorded with a digital oscilloscope (Tektronix, DPO 7104, 1 GHz) with a sampling resistor of 50 Ω or 1 MΩ, respectively. A white LED is used as the bias light source, and a high-speed optical shutter (SR475, time resolution <1 ms) is used to accurately control the on and off of the illumination. After focusing through the lens, the light intensity on the cell can reach to 100 mW/cm$^2$. This optical path can realize a steady-state or low-frequency optical modulation on the cell. An electric circuit consisting of a signal generator (Tektronix, AFG 3052C, 50 MHz) and an LPF is parallel-connected to the cell to give a steady-state or low-frequency electrical modulation. The DC input resistance of this electric circuit is low enough to allow the current from the solar cell produced by the steady-state light illumination or bias voltage to pass through, avoiding the accumulation of free charge and photovoltage. With this design, the bias voltage applied at the cell is only controlled by the signal generator, and the transient signal will not be shunted.

To check the capability and reliability of this system, a commercial multicrystalline silicon solar cell is used as the standard sample for measurement. As shown in Figure 1(b), when the LPF is applied, the recorded transient photovoltage signal is almost the same as that without the electrical modulation circuit. In contrast, if without this LPF, the signal is obviously shunted. Furthermore, the transient photovoltage of the cell under a steady-state light illumination of 100 mW/cm$^2$ and DC bias voltages ranging from 0 to 517 mV are measured, as shown in Figure 1(c), which is reported for the first time, where fluence of the pulse laser is manually adjusted to limit its photovoltage to about 10 mV to ensure small perturbation. Based on the measurements of the standard sample, it is demonstrated that this electro-optical independently modulated transient photocurrent/photovoltage measuring system is reliable.

According to the Poisson equations, the electric field in a solar cell is determined by the charge distribution, while the electric field can be depicted by the perturbation charge transport and recombination, as discussed in the supporting information. Thus, this system can be an effective approach to reveal the interplays between charge and electric field in the perovskite solar cell. In this study, a planar cell with a simple structure of FTO/compact $TiO_2$/perovskite absorber/Spiro-OMeTAD/Au is used, whose scanning electron microscopy image and photovoltaic performance are given in the supporting information. Compared to the mesoscopic one, the planar cell shows much simpler device model, which is more suitable for this study. A series of solar cells has been investigated, and the results discussed in this paper are considered to be universal.

**III. Results and Discussions**

For clarity, the results discussed in the following will be divided into three sections. In the first section, the charge recombination of the perovskite cell under different bias illuminations and voltages is studied with the above electrical transient method. In the second section, the charge collection and differential capacitance of the cell are also experimentally investigated. Further, charge and electric field in the cell are theoretically calculated to explain the experimental results given in the second and third sections and to clarify the interplays between charge and electric field. In the last section, the microscopic charge processes and interplays between charge and electric field behind the photoelectric hysteresis are revealed with both electrical transient experiments and theoretical calculations.

**A. Charge recombination under different bias illuminations and voltages**

The recombination property of the cell is investigated, and the influence of light illumination and bias voltage on the transient photovoltages is shown in Figure 2(a) and (b), respectively. When in dark and under zero bias voltage, the cell exhibits the longest recombination lifetime, beyond one millisecond. However, this lifetime is significantly shortened to several microseconds when the intensity of light illumination increases from 0 to 100 mW/cm$^2$. For comparison, the transient photovoltage of the multicrystalline silicon solar cell has also been measured, as in Figure 2(a), yielding a lifetime of tens of microsecond, no matter in dark or under illumination. Thus, it can be seen that the recombination rate of the perovskite solar cell is very sensitive to light illumination, quite different from the silicon solar cell. Interestingly, no obvious influence of the light illumination on the peak intensity of the photovoltage is found. In contrast, both the peak intensity and photovoltage lifetime obviously decrease when the bias voltage increases from 0 to 1025 mV, as shown in Figure 2(b). More detailed investigation reveals that the recombination rates of the cell under different light illuminations ranging from 0 to 100 mW/cm$^2$ become similar to each other when the bias voltage is higher than 900 mV, as shown in table S1, suggesting that high bias voltages can eliminate the difference caused by light illumination. More discussions about the interplays between charge and electric field will be given in the following.

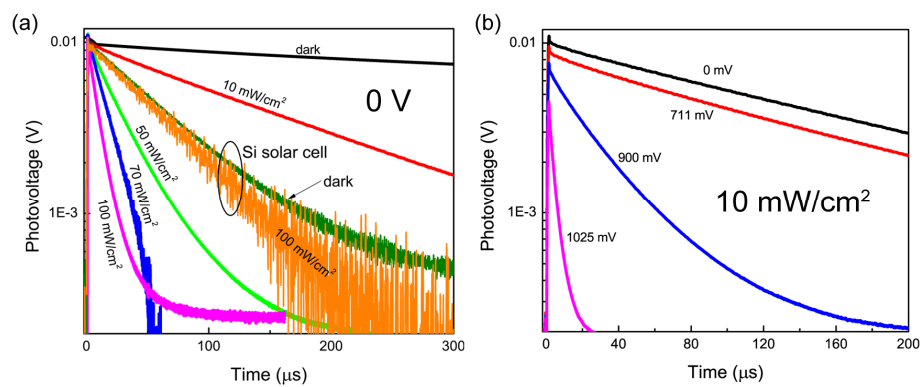

Figure 2. Transient photovoltage results of the perovskite solar cell (a) under bias light illumination with different intensities (0 V) and (b) under different bias DC voltages (10 mW/cm$^2$).

## B. Charge collection and differential capacitance of the cell

The charge collection of the cell under different bias voltages and light illuminations (dark or illumination of 100 mW/cm$^2$) is studied, as shown in Figure 3(a). It is worth noting that the collected charge can be accurately measured with the transient photocurrent method in spite of the influence of cell capacitance. As can be seen, the light illumination has little influence on the charge collection of the cell, which implies that the charge injection under the steady-state light illumination may have no influence on the electric field inside the cell. Moreover, it can also be inferred that the recombination rate does not affect the charge collection, since the photovoltage lifetime of the cell under illumination is almost one hundredth of that in dark, as shown in Figure 2(a). Under bias voltages, the collected charges only show a little decrease when the voltage increases from 0 to 900 mV, but significantly decreases by about 20 times when the voltage further increases from 900 to 1100 mV, indicating that the charge distribution or electric field inside the cell is significantly changed in this stage. Compared to the charge collection, the photovoltage also exhibits a similar behavior depending on the bias voltage, being decreased by about five times when the bias voltage increases from 900 to 1100 mV.

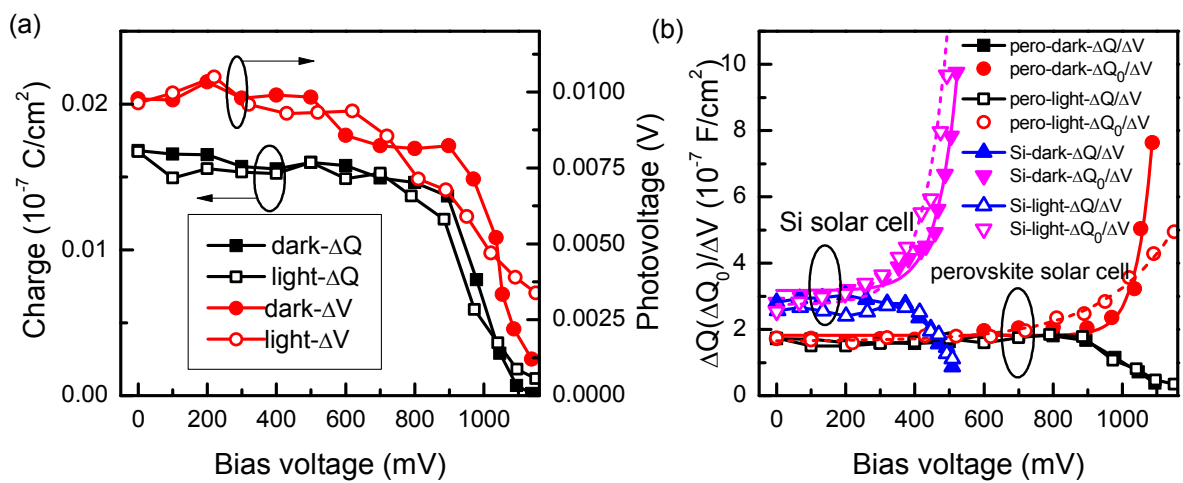

Figure 3. (a) Collected photo induced charge and peak photovoltage, and (b) differential capacitance of the perovskite solar cell under different DC bias voltages in dark and under steady-state light illumination of 100 mW/cm$^2$.

With the collected charge and photovoltage, the differential capacitance of the cell at different bias voltages can be deduced, as shown in Figure 3(b). Here, the differential capacitance is calculated by two different formulas, respectively, which are $C=\Delta Q$ (V)/$\Delta V$ (V) and $C=\Delta Q$ (0 V)/$\Delta V$ (V), the latter was widely used in previous reports [33, 35]. The differential capacitance calculated with the latter one shows an increase when the bias voltage is higher than 900 mV, and can be fitted with an exponential equation.[36] As we know, in the perovskite solar cell, the free carriers have to firstly experience transport and recombination processes in the timescale of nanosecond before collected by the TiO$_2$ layer.[13, 14] Thus, the charge really collected by the selective contacts should be bias voltage dependent, as shown in Figure 3(a). Considering this condition, the differential capacitance is re-calculated as $\Delta Q$ (V)/$\Delta V$ (V), as shown with the black line in Figure 3(b). The capacitance keeps constant with a value of about $1.7\times10^{-7}$ F/cm$^2$ when the bias voltage is lower than 900 mV, close to what has been reported,[33] then decreases gradually to $0.4\times10^{-7}$ F/cm$^2$ when the bias voltage further increases, quite different from previous results [33] and the calculated results with the previous method. For the cell under illumination, the same result is obtained. Interestingly, this behavior can also be observed in the silicon solar cell, where the capacitance gradually decreases when the bias voltage increases from 400 to 500 mV, as shown with the blue line in Figure 3(b). For a solar cell with depletion electric field, the pulse photo-induced carriers would be drifted into and stored in the selective contact layers in the open-circuit condition, and the differential capacitance indicates the charge storage property (e.g. density of states) of the selective contact layers, which can be modeled as a capacitor. However, when the bias voltage is large enough to neutralize the built-in potential ($V_{bi}$), the pulse photo-induced carriers would distribute in both the selective and the absorber layers, which can thus be modeled as two series-connected capacitors. In this case, the equivalent capacitance of the cell is gradually decreased. To check this speculation, Mott-Schottky curves of the perovskite and the silicon solar cells are measured, which gives the $V_{bi}$s of the silicon and perovskite solar cells of about 0.43 V and 0.83 V, respectively. Further considering the conduction band offset in the perovskite solar cell, it is an agreement between the differential capacitance and the Mott-Schottky results. Thus, four inferences can be made from these experimental results. Firstly, the heterojunction depletion electric field plays a critical role in drifting the free carriers into the selective contact layers, and the charge injection produced by steady-state light illumination does not significantly influence the direction or strength of this field or the carrier occupation property in the selective contact layers. Secondly, the bias voltage is mainly applied at the perovskite absorber, and the depletion region mainly locates in the perovskite absorber. Thirdly, free charge can also be stored in the perovskite absorber. Fourthly, the capacitance obtained with this electrical transient method is different from that measured with an impedance spectroscopy method, from both values and physical mechanisms.

To deeply understand the charge recombination, collection and capacitance results, the carrier distribution and electric field inside the cell are theoretically calculated by solving the Poisson equations in dark and under illumination by the general device simulator Analysis of Microelectronic and Photonic Structures (AMPS-1D).[37] It is found that large electrostatic fields of about $10^6$ V/m and $10^5$ V/m exist in the perovskite absorber and the $TiO_2$ interfacial regions, respectively, when no bias voltage is applied no matter in dark or under illumination. When the positive bias voltage is applied and gradually increased, the strength of these fields is obviously decreased. Moreover, it is found that the concentration of free majority carriers in the perovskite absorber will be significantly increased to $10^{12}$~$10^{13}$ cm$^{-3}$ when the light illumination is applied in the short-circuit condition, indicating a co-existence of free charge and high electric field, quite different from the general physics model for solar cells.[38] This high concentration of free carriers at interfaces can well explain the shortened recombination lifetime under illumination. Apart from the light illumination, positive bias voltage can also increase the concentration of free carriers at interfaces. Indeed, the calculation results show that the concentration of free holes at the $TiO_2$/perovskite absorber interfaces can be increased to about $10^{13}$ cm$^{-3}$ when the bias voltage is increased to 900 mV, similar to that under illumination. Thus, the high bias voltage can eliminate the difference in the concentration of free majority carriers at interfaces caused by the illumination, which agrees with our experimental results. Obviously, these results further confirm the significant influence of electric field on the free charge distribution, transport and collection inside the perovskite solar cell.

**C. Microscopic charge processes behind the photoelectric hysteresis**

The microscopic processes regarding charge transport and recombination behind the commonly concerned hysteresis behavior can reflect more interesting and complicated interplays between charge and electric field in this cell. The hysteresis behavior in the perovskite solar cell means that the current-voltage (*I-V*) characteristics and output performance of the cell is significantly dependent on the measuring conditions, such as scanning direction and rate, pre-bias conditions.[23-25, 39-44] In our previous work, we observed an intrinsic slow charge response in the photocurrent measurements and attributed it to ion migration.[34] Thus, all the interplays among static ion charge, free charge and the electric field should be involved in this process. Compared to transient photocurrent, transient photovoltage, especially its lifetime, can give more information on the strength of electric field, carrier distribution and defect properties inside the cell.

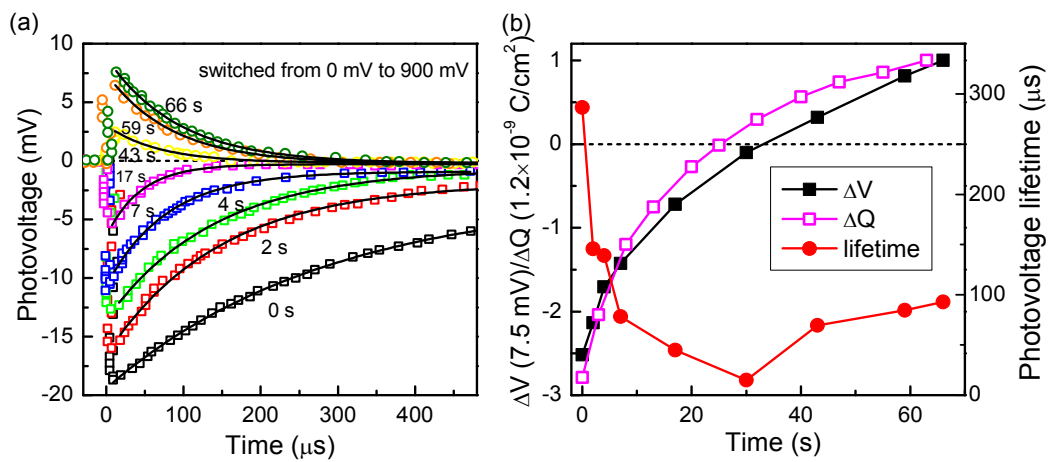

Figure 4. (a) Hysteresis behavior of the transient photovoltage just after the bias voltage at the cell is switched from 0 to 900 mV and (b) the evolution of collected photo-induced charge, peak photovoltage and the photovoltage lifetime with time in dark.

Figure 4(a) shows the transient photovoltage results just after the bias voltage at the perovskite solar cell is switched from 0 to 900 mV in dark, exhibiting an obvious hysteresis process. In this process, a negative photovoltage signal is firstly observed with a long lifetime of about 290 μs. After that, the absolute value of the photovoltage gradually decreases, accompanied by the decrease in recombination lifetime. After about 30 s, the positive signal appears with a very short lifetime of about several microseconds, and then gradually increases in both the intensity and lifetime. Correspondingly, the collected charge also exhibits a similar hysteresis process from negative to positive, as shown in Figure 4(b). Thus, the hysteresis in the intensity of photovoltage may result from the decay process in charge collection. In the general carrier transport model for a semiconductor device as discussed in the supporting information, the transport property of the non-equilibrium charge is mainly determined by the electric field and boundary conditions. Since the boundary conditions, such as band offset, charge extraction rate, may not be significantly influenced, this decay process in charge collection implies a hysteresis in the direction and strength of the internal electric field inside the cell, especially inside the perovskite absorber. As has been discussed, the ion (vacancy) migration is the most possible origin for such a hysteresis in internal electric field.[34, 45-46] Theoretical simulations also support that the doping effect and accumulation of ion charge caused by ion migration can indeed weaken and even reverse the built-in electric field inside the perovskite absorber (Figure S5), which agrees well with the experimental results. This hysteresis in the electric field can also explain the evolution of photovoltage lifetime, since a large electric field can effectively separate free electrons and holes, suppressing recombination. Recombination increases gradually when strength of the negative electric field decreases, whereas decreases when strength of the positive electric field increases.

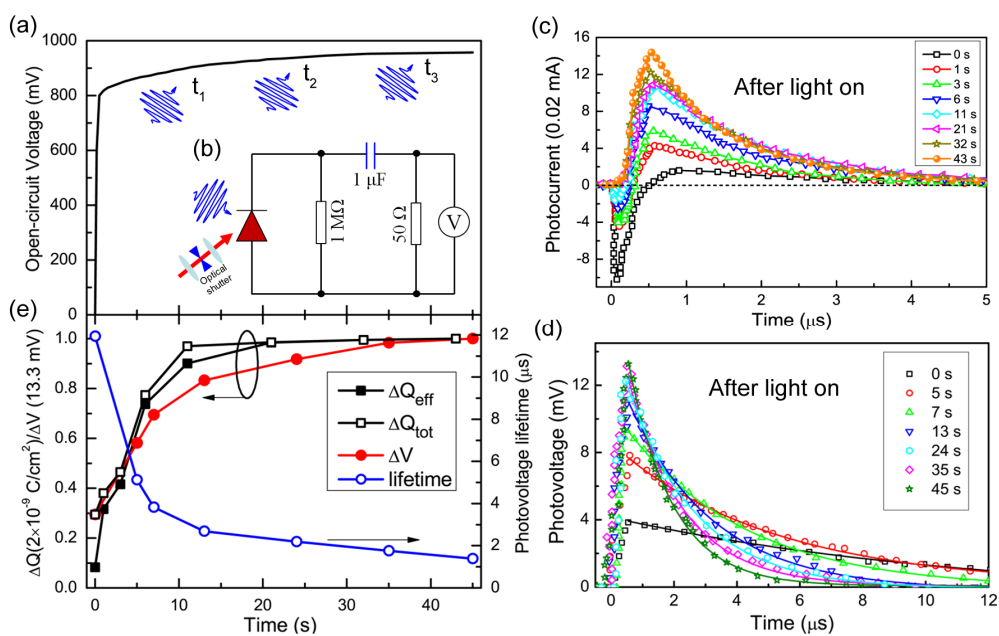

Figure 5. (a) Hysteresis behavior of the open-circuit voltage ($V_{OC}$) of the perovskite solar cell under the steady-state light illumination of 100 mW/cm$^2$, (b) schematic diagram of the designed electric circuit for probing the transient photocurrent in the open-circuit condition, (c) transient photocurrent and (d) photovoltage results during the $V_{OC}$ hysteresis, and (e) evolution of collected photo induced charge, peak photovoltage and photovoltage lifetime with time during the $V_{OC}$ hysteresis.

Hysteresis behavior of the open-circuit voltage ($V_{OC}$) in this cell was also observed.[46-48] As shown in Figure 5(a), when the light illumination is switched on rapidly within 1 ms, a low $V_{OC}$ of about 800 mV appears firstly, and then increases gradually to about 960 mV, taking a long time of more than 50 s. In comparison, $V_{OC}$ of the standard silicon solar cell can increase to a stable value rapidly within the resolution time of the optical shutter. Thus, revealing the charge processes behind this behavior is another approach to further understand the photoelectric characteristics of this cell. To realize the probing of charge transport of the cell working in the open-circuit condition, an HPF electric circuit is designed, as depicted in Figure 5(b). Our design gives a direct approach to acquire the complete transient photocurrent signal of the cell, thus being able to probe more complicated photoelectric processes, in comparison to the known method [30]. Figure 5(c) shows the transient photocurrents of the cell which is experiencing the hysteresis in $V_{OC}$ after turning on the light illumination. At the time of 0 s, a sharp negative photocurrent signal with an intensity of about -0.2 mA appears firstly, and then increases to positive, showing a peak photocurrent of about 0.04 mA. At the time of 1, 3, 6, 11 s, this sharp negative signal can always be observed, but its absolute intensity gradually decreases. After a time of about 20 s, this negative signal disappears. The peak intensity of the positive part of the transient photocurrent signal increases gradually at the same time. This means that a proportion of free electrons (or holes) excited by the pulse laser are drifted or diffused toward the hole (or electron) transport layer while another proportion transport in the opposite direction at the same time in the early stage after the cell being illuminated. Due to the difference in carrier transport velocity through the front or back contact regions, the negative photocurrent signal appears at first. In the later stage of the $V_{OC}$ hysteresis, most of the free electrons can transport toward and be collected by the electron transport layer, showing positive photocurrent signals. Besides the photocurrent, the transient photovoltage results exhibit a more interesting hysteresis behavior, as shown in Figure 5(d). At the time of 0 s, the photovoltage shows a low peak value of about 4 mV, but decay in a slow rate, yielding a recombination lifetime of about 12 μs. After that, the peak intensity increases gradually before saturates while the recombination lifetime decreases continuously to about 1.3 μs. This hysteresis process in photovoltage and its lifetime is quite different from the results shown in Figure 4, implying a more complicated charge processes inside the cell during the $V_{OC}$ hysteresis. Figure 5(e) summarizes the evolutions of collected photo-induced charge ($\Delta Q$), peak photovoltage ($\Delta V$) and photovoltage lifetime with time during the $V_{OC}$ hysteresis, where $\Delta Q_{tot}$ means the total collected charge as $|\Delta Q_{negative}|+|\Delta Q_{positive}|$, and $\Delta Q_{eff}$ means the effective charge collected by the selective contacts as $\Delta Q_{negative}+\Delta Q_{positive}$, in which $\Delta Q_{negative}$ and $\Delta Q_{positive}$ are obtained by integrating the negative and positive part of the transient photocurrent, respectively. The $\Delta V$ shows a similar evolution process to that of the $\Delta Q$, implying that the hysteresis in the peak intensity of photovoltage is mainly due to the hysteresis in effective charge collection. This hysteresis in charge collection can be similarly explained by the hysteresis in electric field inside the cell, as discussed above. However, this simple process can hardly explain the hysteresis in photovoltage lifetime, which decreases obviously when the photovoltage increases. Moreover, whether this simple process about electric field can explain the decay in $V_{OC}$ is still a question.

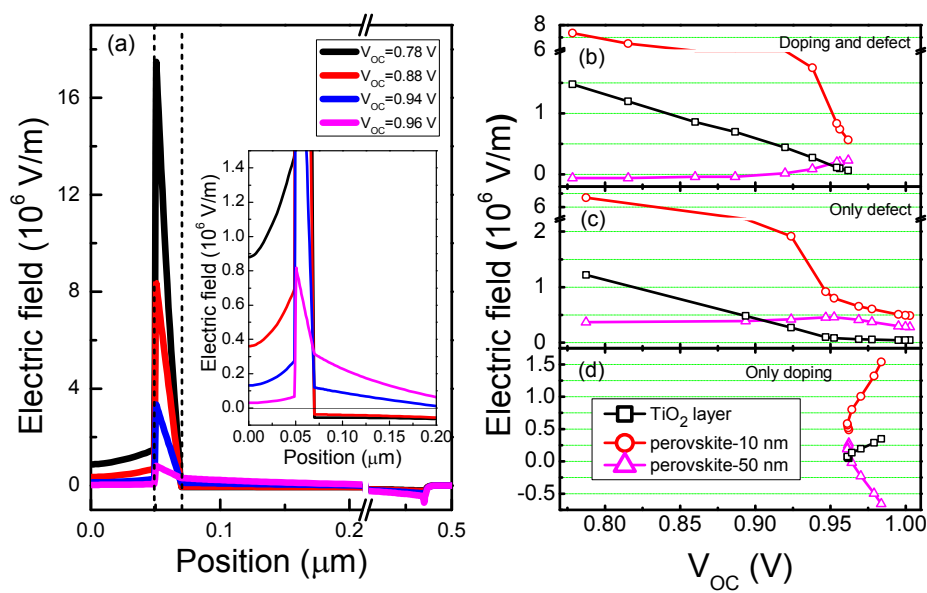

Figure 6. (a) The electric field inside the cell with different accepter and defect density in the perovskite interfacial region with a width of 20 nm close to the $TiO_2$/perovskite interface, yielding different $V_{OC}$. The relationship between the strength of electric field in the $TiO_2$ layer ($x$=49 nm), perovskite interfacial region ($x$=60 nm) and perovskite absorber ($x$=100 nm) with the $V_{OC}$ of the cell when considering the influence of ion migration on (b) both doping and defect density , (c) only defect density and (d) only doping density.

Theoretical calculations of the charge and electric field properties of the cell under light illumination are carried out, and an interfacial perovskite region with a width of 20 nm (close to TiO$_2$/perovskite interface) with different accepter and defect density coming from ion (vacancy) migration is considered here as origins of the hysteresis.[49, 50] The cells with different device and material parameters (doping and defect density) are simulated, details are given in the supporting information. The theoretical results are compared with our experimental results to understand what occurs during the hysteresis process. According to previous discussions, the vacancies of CH$_3$NH$_3$ and Pb, interstitial and substitutional iodide may be drifted toward and located in the interfacial region under the electrostatic forces coming from the built-in electric field,[34, 45-46] causing the appearance and enhancement of shallow accepter doping and deep defect.[51] Figure 6(a) gives the calculation results of the electric field inside the cells with different accepter and defect density in the interfacial region, yielding different $V_{OC}$ ranging from 0.78 to 0.96 V. As shown in the inset, the strength and even the direction of electric fields in the TiO$_2$/perovskite regions of the cells with different $V_{OC}$ are quite different from each other. The relationship between the strength of electric field in the TiO$_2$ layer ($x$=49 nm), perovskite interfacial region ($x$=60 nm) and perovskite absorber ($x$=100 nm) derived from Figure 6(a) with the $V_{OC}$ of the cell is shown in Figure 6(b). As can be seen, the strength of electric field both in the TiO$_2$ and perovskite interfacial layer is obviously increased when the $V_{OC}$ of the cell is decreased from 0.96 to 0.78 V. However, the strength of electric field in the perovskite absorber decreases instead, from positive to negative, indicating a reversed direction of the electric field. This theoretical result about the electric fields inside the cell can well explain the experimental behaviors of transient photocurrent/photovoltage we observed during the $V_{OC}$ hysteresis process.

In the early stage of this hysteresis, the perovskite interfacial region has a high concentration of shallow accepter doping and deep defect due to ion (vacancy) accumulation, yielding a low $V_{OC}$. Due to the negative electric field inside the perovskite absorber, a large proportion of photo-induced free electrons (or holes) is drifted and diffused toward the hole (or electron) transport layer and then the external electric circuit, yielding a sharp negative photocurrent signal. During the transport, numerous of free carriers are also recombined due to the existence of high concentration of deep defects, leading to a low $\Delta Q_{tot}$. In the photovoltage measurement, the collected carriers are stored in the selective contacts and the electrodes, far away from the interfaces due to the high positive electric field in $TiO_2$, thus resulting in the slow recombination. Afterwards, the concentration of accepter and defect in the interfacial region decreases gradually due to the ion (vacancy) remigration, [34] thus yielding a higher $V_{OC}$ and obvious change in the electric field. Correspondingly, more free electrons can be collected by the electron transport layer and fewer carriers are recombined during transport, leading to higher $\Delta Q_{eff}$ and $\Delta Q_{tot}$. On the other hand, the recombination is instead enhanced due to the decrease in the strength of electric fields in the $TiO_2$ and perovskite interfacial regions. In the later stage of this hysteresis, a new balance between ion (vacancy) migration and the electric field is established. In this case, most free carriers excited by the pulse laser can be collected, but exhibit short photovoltage lifetimes.

For comparison, the single effect of doping or defect is also considered. The relationship between electric fields inside the cell with $V_{OC}$ is shown in Figure 6(c) when only the influence of ion accumulation on the defect density is considered. Although the relationship between strength of electric fields in the $TiO_2$ and perovskite interfacial regions with the $V_{OC}$ is similar to that shown in Figure 6(b), the electric field in the perovskite absorber of the cell with a $V_{OC}$ of about 0.78 V is even stronger than that of the cell with $V_{OC}$ of about 1 V, which cannot explain the transient photocurrent results we observed. When only the doping effect is considered, the cells with different accepter density in the interfacial region all show high $V_{OC}$s, which cannot explain the $V_{OC}$ hysteresis. Moreover, the increase in accepter density in the interfacial region can even increase the $V_{OC}$, which is different from the conventional inorganic junction cell [52] and might be the distinct characteristic of the perovskite solar cell scaled in the sub-micrometer. Therefore, the hysteresis process occurred in the open-circuit condition is suggested to be attributed to the effect of ion migration on both the doping and defect properties of the perovskite absorber. Furthermore, the significant influence of static charges derived from doping and defect on the electric field as well as on the charge transport and recombination in the cell is also demonstrated.

## IV. Conclusions

Important interplays between charge and electric field, which play a critical role in determining the charge transport, recombination, storage and hysteresis in the perovskite solar cell, have been systematically investigated by the modulated electrical transient methods and theoretical calculations. It is found that high concentration of free carriers and strong electric field can co-exist inside the perovskite absorber, and that the charge transport and collection capacity of the cell is dependent on the electric field. Interestingly, the high concentration of free carriers produced by the steady-state illumination does not obviously influence the electric field in the cell, but significantly increase the recombination rate of non-equilibrium charge. The charge storage property of this cell is very similar to that of the silicon solar cell, indicating a similar charge and junction mechanism and that the electric field determines the distribution of free charge. Moreover, it is found that the static charge coming from both doping and defect can significantly influence the electric field inside the cell, thus affecting charge processes, further leading to the appearance of hysteresis. We believe these clarifications of the fundamental working mechanisms of the cell can give a guide to solve the hysteresis and stability problems with physical and chemical engineerings of the doping and defect at interfaces, and that this electrical transient method can be an effective and universal approach for the investigation of charge processes in related photoelectric-physics and chemistry systems.

**V. Experimental Section**

*Cell fabrication.* Firstly, a dense $TiO_2$ compact layer with a thickness of about 50 nm was spin-coated onto the pre-cleaned laser patterned FTO glass, then sintered at 500 °C for 30 min. The (FA, MA)$PbI_3$ absorber film was deposited onto the $TiO_2$ layer with a repeated interdiffusion method. Firstly, a 1.3 M $PbI_2$ solution was spin coated at a speed of 4000 rpm and heated on a hot plate at 70 °C for 2 min, forming a transparent yellow film. After cooling down to room temperature, the (FA, MA)I isopropanol solution (mole ratio FA: MA=2:1) with a concentration of 30 mg/ml was repeatedly spin coated onto the $PbI_2$ film to give a smooth film.[53] This film was then rinsed with isopropanol to remove the (FA, MA)I residual, heated at 120 °C for 100 min to completely transform to a transparent dark brown perovskite film. Spiro-OMeTAD was then spin coated onto the perovskite film as a hole transport layer (HTL). This FTO/$TiO_2$/perovskite/HTL film was kept in dark in air (humidity: 20%) overnight. Finally, Au was thermally evaporated (Kurt J. Lesker) as back electrode (80 nm) at an atmospheric pressure of $10^{-7}$ Torr.

*Characterization of the cell.* Current-Voltage (*I-V*) characteristics were measured by an additional voltage from the Keithley 2602 system source meter together with a sunlight simulator (Newport 91160A, 100 mW/cm$^2$) calibrated with a standard silicon reference cell. The solar cells were masked with a black aperture to define the active area of 0.1 cm$^2$. Before *I-V* measuring, no bias voltage and illumination was pre-applied at the cell. The current was recorded every 10 mV and delayed for 0.1 or 0.5 s. Scanning electron microscopy (SEM) images were obtained with FEI-SEM (XL 30 S-FEG).

For the electrical transient measurements, the non-equilibrium carriers was excited by a 532 nm laser (Brio, 20 Hz) with the duration of 4 ns and pulse fluence of about 5 nJ/cm$^2$ (per pulse). A digital oscilloscope (Tektronix, DPO 7104) was used to record the photocurrent or photovoltage decay process with a sampling resistance of 50 Ω or 1 MΩ, respectively. A digital signal generator (Tektronix, AFG 3052C, 50 MHz) was used to give an external modulation at the cell. The signal generator was AC separated from the cell with a low-pass filter. A white LED with a light intensity of 100 mW/cm$^2$ was used as the bias illumination, and a high-speed optical shutter was used to control the on and off of the illumination. For all the measurements, the transient signals were begun to be recorded five minutes after the external modulations were applied and averaged to get a reliable result.

A series of perovskite solar cells were investigated, and the results discussed in this paper are considered to be universal. The cycle time of the pulse laser is about 50 ms, much larger than the transient time of the cell; and the interference between different transient photocurrent signals is thus negligible.

*Theoretical calculations of the photoelectric characteristics of the cell.* The carrier distribution and electric field inside the cell is theoretically calculated by solving the Poisson equations and charge conservation and continuity equations in dark and under illumination by the general device simulator Analysis of Microelectronic and Photonic Structures (AMPS-1D). The detailed material and device parameters are given in the supporting information.

**Acknowledgements**

This work was supported by Natural Science Foundation of China (Grant Nos. 51372270, 11474333, 91233202 and 91433205).**References**

1. A. Kojima, K. Teshima, Y. Shirai, T. Miyasaka, *J. Am. Chem. Soc.* **131**, 6050 (2009).

2. H. S. Kim, C. R. Lee, J. H. Im, K. B. Lee, T. Moehl, A. Marchioro, S. J. Moon, R. H. Baker, J. H. Yum, J. E. Moser, M. Grätzel, N. G. Park, *Sci. Rep. 2*, 591 (2012).

3. M. M. Lee, J. Teuscher, T. Miyasaka, T. N. Murakami, H. J. Snaith, *Science 338*, 643 (2012).

4. J. Burschka, N. Pellet, S. J. Moon, R. H. Baker, P. Gao, M. K. Nazeeruddin, M. Grätzel, *Nature 499*, 316 (2013).

5. M. Liu, M. B. Johnston, H. J. Snaith, *Nature 501*, 395 (2013).

6. J. Xiao, J. Shi, D. M. Li, Q. B. Meng, *Sci. China Chem. 58*, 221 (2015).

7. N. J. Jeon, J. H. Noh, W. S. Yang, Y. C. Kim, S. Ryu, J. Seo, S. Il Seok, *Nature 517*, 476 (2015).

8. H. Zhou, Q. Chen, G. Li, S. Luo, T. Song, H. -S. Duan, Z. Hong, J. You, Y. Liu, Y. Yang, *Science 345*, 542 (2014).

9. W. S. Yang, J. H. Noh, N. J. Jeon, Y. C. Kim, S. Ryu, J. Seo, S. Il Seok, *Science 348*, 1234 (2015).

10. D. Bi, W. Tress, M. I. Dar, P. Gao, J. Luo, C. Renevier, K. Schenk, A. Abate, F. Giordano, J. -P. C. Baena, J. -D. Decoppet, S. M. Zakeeruddin, M. K. Nazeeruddin, M. Grätzel, A. Hagfeldt, *Sci. Adv.* **2016**, doi: 10.1126/sciadv.1501170.

11. T. Zhang, Y. Zhao, *Acta Chim. Sinica 73*, 202 (2015).

12. H. L. Yuan, J. P. Li, M. K. Wang, *Acta Phys. Sin. 64*, 038405 (2015).

13. G. Xing, N. Mathews, S. Sun, S. Lim, Y. M. Lam, M. Grätzel, S. Mhaisalkar, T. C. Sum, *Science 342*, 344 (2013).

14. S. D. Stranks, G. E. Eperon, G. Grancini, C. Menelaou, M. J. Alcocer, T. Leijtens, L. M. Herz, A. Petrozza, H. J. Snaith, *Science 342*, 341 (2013).

15. A. Marchioro, J. Teuscher, D. Friedrich, M. Kunst, R. Krol, T. Moehl, M. Grätzel, J. E.18

Moser, *Nat. Photon. 8*, 250 (2014).

16. Jr. C. S. Ponseca, T. J. Savenije, M. Abdellah, K. Zheng, A. Yartsev, T. Pascher, T. Harlang, P. Chabera, T. Pullerits, A. Stepanov, J. P. Wolf, V. Sundstrom, *J. Am. Chem. Soc. 136*, 5189 (2014).

17. H. S. Kim, Mora-Sero Ivan, V.Gonzalez-Pedro, F F.abregat-Santiago, E. J. Juarez-Perez, N. G. Park, J. Bisquert, *Nat. Commun. 4*, 2242 (2013).

18. M.Bag, L. A. Renna, R. Y. Adhikari, S. Karak, F. Liu, P. M. Lahti, T. P. Russell, M. T. Tuominen, D. Venkataraman, *J. Am. Chem. Soc. 137*, 13130 (2015).

19. A. Dualeh, T. Moehl, N. Tetreault, J. Teuscher, P. Gao, M. K. Nazeeruddin, M. Grätzel, *ACS Nano 8*, 362 (2014).

20. J. Shi, X. Xu, D. M. Li, Q. B. Meng, *Small 11*, 2472 (2015).

21. E. Edri, S. Kirmayer, S. Mukhopadhyay, K. Gartsman, G. Hodes, D. Cahen, *Nat. Commun. 5*, 3461 (2014).

22. V. W. Bergmann, S. A. L. Weber, F. J. Ramos, M. K. Nazeeruddin, M. Grätzel, D. Li, A. L. Domanski, I. Lieberwirth, S. Ahmad, R. Berger, *Nat. Commun. 5*, 5001 (2014).

23. H. J. Snaith, A. Abate, J. M. Ball, G. E. Eperon, T. Leijtens, N. K. Noel, S. D. Stranks, J. T.-W. Wang, K. Wojciechowski, W. Zhang, *J. Phys. Chem. Lett. 5*, 1511 (2014).

24. E. L. Unger, E. T. Hoke, C. D. Bailie, W. H. Nguyen, A. R. Bowring, T. Heumuller, M. G. Christoforo, M. D. McGehee, *Energy Environ. Sci. 7*, 3690 (2014).

25. H. S. Kim, N. G. Park, *J. Phys. Chem. Lett. 5*, 2927 (2014).

26. W. J. Yin, T. Shi, Y. Yan, *Adv. Mater. 26*, 4653 (2014).

27. Y. Yuan, J. Chae, Y. Shao, Q. Wang, Z. Xiao, A. Centrone, J. Huang, *Adv. Energy Mater. 5*, 1500615 (2015).

28. T. Leijtens, A. R. S. Kandada, G. E. Eperon, V. D'Innocenzo, J. M. Ball, S. D. Stranks, H. J. Snaith, A. Petrozza, *J. Am. Chem. Soc. 137*, 15451 (2015).

29. P. R. F. Barnes, K. Miettunen, X. Li, A. Y. Anderson, T. Bessho, M. Grätzel, B. C. O'Regan, *Adv. Mater. 25*, 1881 (2013).

30. B. C. O'Regan, K. Bakker, J. Kroeze, H. Smit, P. Sommeling, J. R. Durrant, *J. Phys. Chem. B 110*, 17155 (2006).

31. F. Deledalle, T. Kirchartz, M. S. Vezie, M. Campoy-Quiles, P. S. Tuladhar, J. Nelson, J.

R. Durrant, *Phys. Rev. X* 5, 011032 (2015).

32. F. Marlow, A. Hullermann, L. Messmer, *Adv. Mater. 27*, 2447 (2015).

33. B. C. O'Regan, P. R. F. Barnes, X. Li, C. Law, E. Palomares, J. M. Martin-Beloqui, *J. Am. Chem. Soc. 137*, 5087 (2015).

34. J. Shi, X. Xu, H. Zhang, Y. H. Luo, D. M. Li, Q. B. Meng, *Appl. Phys. Lett. 107*, 163901 (2015).

35. K. Wojciechowski, M. Saliba, T. Leijtens, A. Abate, H. J. Snaith, *Energy Environ. Sci. 7*, 1142 (2014).

36. J. Bisquert, *Phys. Chem. Chem. Phys. 13*, 4679 (2011).

37. H. Zhu, A. K. Kalkan, J. Y. Hou, S. J. Fonash, *AIP Conf. Proc. 462*, 309 (1999).

38. A. Luque, S. Hegedus, Handbook of photovoltaic science and engineering. John Wiley & Sons, UK 2011.

39. W. Tress, N. Marinova, T. Moehl, S. M. Zakeeruddin, M. K. Nazeerudin, M. Grätzel, *Energy Environ. Sci. 8*, 995 (2015).

40. J. Wei, Y. Zhao, H. Li, G. Li, J. Pan, D. Xu, Q. Zhao, D. Yu, *J. Phys. Chem. Lett. 5*, 3937 (2014).

41. A. K. Jena, H. W. Chen, A. Kogo, Y. Sanehira, M. Ikegami, T. Miyasaka, *ACS Appl. Mater. Interfaces 7*, 9817 (2015).

42. Y. Shao, Z. Xiao, C. Bi, Y. Yuan, J. Huang, *Nat. Commun. 5*, 5784 (2014).

43. H. Zhang, X. Xiao, Y. Shen, M. Wang, J. Energy Chem. 24, 729 (2015).

44. Y. Zhang, Z. Yao, S. Lin, J. Li, H. Lin, *Acta Chim. Sinica 73*, 219 (2015).

45. Z. Xiao, Y. Yuan, Y. Shao, Q. Wang, Q. Dong, C. Bi, P. Sharma, A. Gruverman, J. Huang, *Nat. Mater. 14*, 193 (2015).

46. Y. Deng, Z. Xiao, J. Huang, *Adv. Energy Mater. 5*, 1500721 (2015).

47. S. D. Stranks, V. M. Burlakov, T. Leijtens, J. M. Ball, A. Goriely, H. J. Snaith, *Phys. Rev. Appl. 2*, 034007 (2014).

48. C. Zhao, B. Chen, X. Qiao, L. Luan, K. Lu, B. Hu, *Adv. Energy Mater. 5*, 1500279 (2015).

49. O. Almora, I. Zarazua, E. Mas-Marza, I. Mora-Sero, J. Bisquert, G. Garcia-Belmonte, *J. Phys. Chem. Lett. 6*, 1645 (2015).

50. S. van Reenen, M. Kemerink, H. J. Snaith, *J. Phys. Chem. Lett. 6*, 3808 (2015).

51. W. J. Yin, T. Shi, Y. Yan, *Appl. Phys. Lett. 104*, 063903 (2014).

52. R. Scheer, H. W. Schock, Chalcogenide Photovoltaics, Wiley-VCH, Germany 2011.

53. H. Zhang, J. Shi, J. Dong, X. Xu, Y. H. Luo, D. M. Li, Q. B. Meng, *J. Energy Chem. 24*, 707 (2015).